\documentclass[manuscript]{aastex}
\usepackage{amsmath}

\shorttitle{NLTE SPS of GC w/ S-ILS I}
\shortauthors{Young \& Short}

\begin{document}

%% LaTeX will automatically break titles if they run longer than
%% one line. However, you may use \\ to force a line break if
%% you desire.

\title{NLTE Stellar Population Synthesis of Globular Clusters using Synthetic Integrated Light Spectra I: Constructing the IL Spectra}

%% Use \author, \affil, and the \and command to format
%% author and affiliation information.
%% Note that \email has replaced the old \authoremail command
%% from AASTeX v4.0. You can use \email to mark an email address
%% anywhere in the paper, not just in the front matter.
%% As in the title, use \\ to force line breaks.

\author{Mitchell. E. Young}
\affil{Department of Astronomy \& Physics and Institute for Computational Astrophysics, Saint Mary's University,
    Halifax, NS, Canada, B3H 3C3}
\email{myoung@ap.smu.ca}

\author{C. Ian Short}
\affil{Department of Astronomy \& Physics and Institute for Computational Astrophysics, Saint Mary's University,
    Halifax, NS, Canada, B3H 3C3}
\email{}

%% Mark off your abstract in the ``abstract'' environment. In the manuscript
%% style, abstract will output a Received/Accepted line after the
%% title and affiliation information. No date will appear since the author
%% does not have this information. The dates will be filled in by the
%% editorial office after submission.

\begin{abstract}

We present an investigation of the globular cluster population synthesis method 
of \citet{mcwilliam08}, focusing on the impact of NLTE modeling effects and CMD 
discretization. Johnson-Cousins-Bessel U-B, B-V, V-I, and J-K colors are produced for 96 
synthetic integrated light spectra with two different discretization prescriptions 
and three degrees of NLTE treatment. These color values are used to compare NLTE 
and LTE derived population ages. Relative contributions of different spectral 
types to the integrated light spectra for different wavebands are measured.
Integrated light NLTE spectra are shown to be more luminous in the UV and optical 
than LTE spectra, but show stronger absorption features in the IR. 
The main features showing discrepancies between NLTE and LTE integrated light 
spectra may be attributed to light metals, primarily Fe I, Ca I, and Ti I, as well 
as TiO molecular bands. 
Main Sequence stars are shown to have negligible NLTE effects at IR wavelengths 
compared to more evolved stars.
Photometric color values are shown to vary at the millimagnitude level as a function of 
CMD discretization. Finer CMD sampling for the upper main sequence and turnoff, 
base of the red giant branch, and the horizontal branch minimizes this variation.
Differences in ages derived from LTE and NLTE 
IL spectra are found to range from 0.55 to 2.54 Gyr, comparable to the uncertainty 
in GC ages derived from color indices with observational uncertainties of 0.01 
magnitudes, the limiting precision of the Harris catalog \citep{harris96}.

\end{abstract}
%% Keywords should appear after the \end{abstract} command. The uncommented
%% example has been keyed in ApJ style. See the instructions to authors
%% for the journal to which you are submitting your paper to determine
%% what keyword punctuation is appropriate.


\section{INTRODUCTION}

Because of their old ages, relative 
homogeneity of their populations, and high luminosities, globular clusters (GCs) can be used to study
the chemical evolution history of galaxies. The most luminous 
GCs presumably only trace the major star forming events, including mergers. Even 
basic metallicity provides interesting information for comparison with the Galaxy. 
However, the detailed chemical composition of GCs could potentially provide much more
information on galaxy evolution, because the chemical elements are produced by 
a variety of stars, with varying sensitivity to stellar mass and metallicity.  
High resolution spectroscopic chemical abundance analysis of individual stars in 
Galactic GCs has long been pursued \citep{cohen78,pilachowski80}.
Such abundance studies are a useful tool for probing the chemical evolution of 
the Galaxy \citep{sneden91,briley94}. Unfortunately, similar studies 
have not been possible outside of the Milky Way, 
as individual stars cannot be resolved in distant galaxies. 

\paragraph{}

For extragalactic GCs, cluster metallicities have been estimated using broadband 
photometric colors \citep{forte81,geisler96} and low-resolution 
integrated light (IL) spectra \citep{racine78,brodie90}. 
Results from such studies include the discovery of bimodal GC metallicity and color
distributions in extragalactic systems \citep{elson96,whitmore95} 
reminiscent of the bimodal GCs in the Milky Way \citep{zinn85}.

\paragraph{}

In recent years, stellar population synthesis (SPS) of GC populations has provided 
a new avenue of investigating the chemical composition of either spatially 
resolved or unresolved GCs, provided high spectral-resolution IL 
spectra can be obtained \citep{mcwilliam08,colucci09,colucci11}. Using both broadband photometric colors
and equivalent widths (EWs) of spectroscopic absorption lines as diagnostics, detailed information 
on the chemical composition of a cluster as a whole can be derived.

\paragraph{}

Non-local thermodynamic equilibrium effects have been demonstrated to be present 
in the spectra of asymptotic-giant branch (AGB) stars in the Galactic GCs 47 Tuc 
\citep{lapenna14}, affecting the analysis of Fe I and II abundances. While the same effects 
have yet to be demonstrated in red-giant branch (RGB) GC stars, there is a possibility that 
the effects will be reflected in the IL spectra of a cluster if it is present in 
its brightest members. 

\subsection{Present Work}

Our primary goal is to investigate NLTE modeling effects on the IL spectra of 
synthetic GCs, as well as ages and metallicities subsequently derived from them,
using the population synthesis methodology presented by \citet{mcwilliam08}. 
We test two degrees of NLTE treatment, one where we model only the 
component of the population more evolved than the sub-giant branch in NLTE, and 
the other where we model the entire population in NLTE. WE also investigate the 
uncertainty associated with the color-magnitude diagram (CMD) discretization used in this method.
We compare ages for 
synthetic clusters derived from photometric colors of both LTE and NLTE spectra, 
and we assess the significance of deviations caused by NLTE effects by comparing 
them to those caused by photometric uncertainty.

\section{LIBRARY OF SYNTHETIC SPECTRA \label{spectra}}

\begin{figure}
\plotone{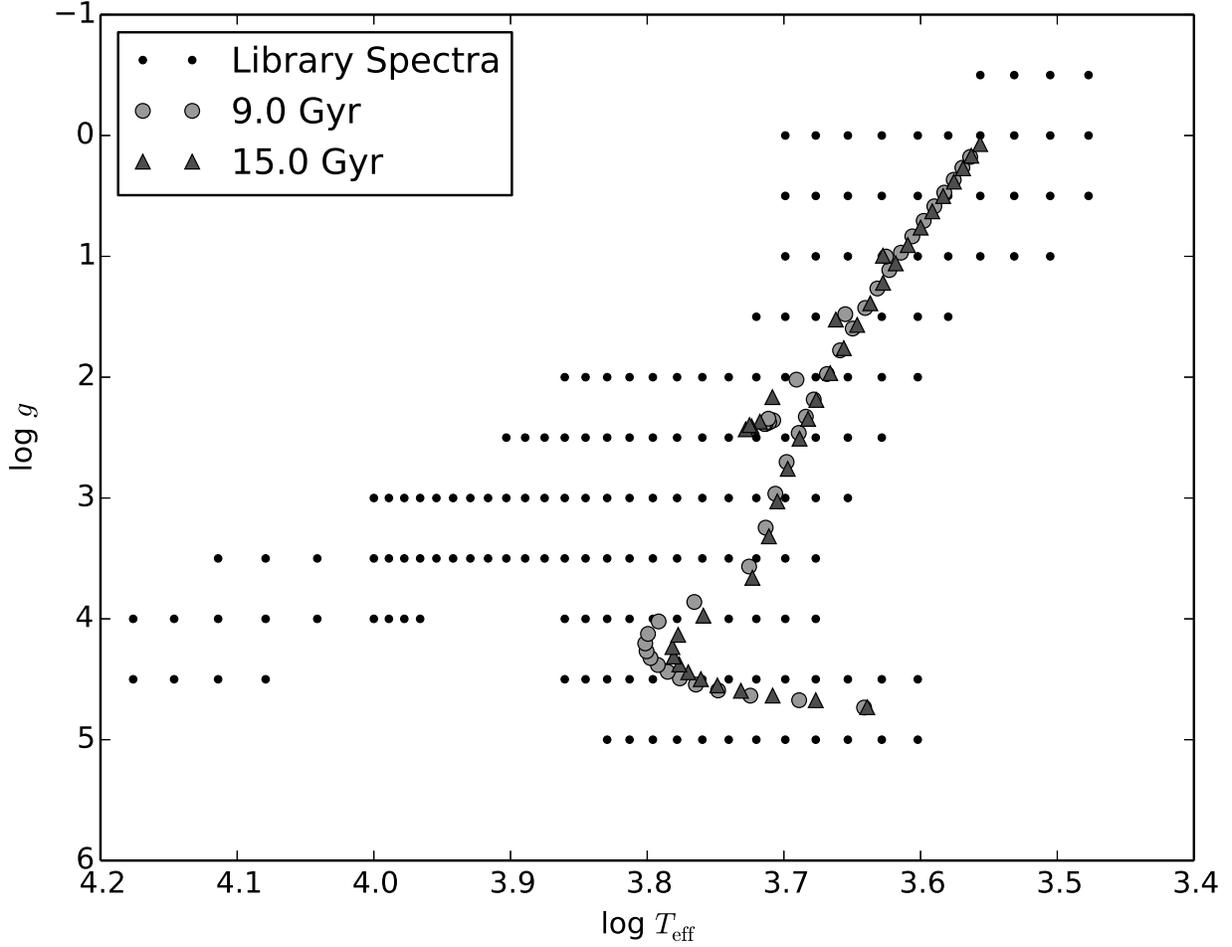}
\caption{Coverage of our library of stellar atmospheres and spectra in $T_\mathrm{eff}$ vs log $g$ space.  This selection is reproduced at both 0.5 and 1 $M_\odot$ and at each of [M/H] = $-0.66$, $-1.0$, and $-1.49$. Two sample representative populations produced from the 9.0 and 15.0 Gyr isochrones have been plotted to indicate where library interpolation occurs. \label{fig:library}}
\end{figure}

We used PHOENIX v.15 to compute spherically symmetric model atmospheres and 
high resolution synthetic spectra (R $\approx$ 300000) for stars that cover the CMD
parameter spaces of GCs at various ages and metallicities spanning Galactic values. 
We produced a library of stellar atmospheres covering the ranges of $T_{\mathrm{eff}}$ 
= 3000 to 15000 K (in steps of 200 K below $T_{\mathrm{eff}}$ = 4000 K, 1000 K 
above $T_{\mathrm{eff}}$ = 10000 K, and 250 K otherwise) and -0.5 to 5.0 dex in 
log $g$ (in steps of 0.5 dex) \citep{coelho14}.  Figure \ref{fig:library} shows 
the extent of the coverage in $T_{\mathrm{eff}}$ vs log $g$ space.
This coverage was reproduced for three values of metallicity, [M/H] = $-1.49$, 
$-1.00$, and $-0.66$, and two values of stellar mass, M = 0.5 and 1.0 $\mathrm{M_\odot}$.  

\paragraph{}

Our library was built in two halves, ``warm'' stars and ``cool'' stars, using a separate 
pipeline for each. For cool star models, we consider 47 different molecules, with 
a combined total of 119 isotopologues and isotopomers, in both the equation of state (E.O.S.)
and opacity calculations. The molecules taken into consideration are listed in Table 
\ref{tab:molecules list}. The atmospheres are also left to naturally form convection 
zones. For warm star models, we do not consider molecules in the E.O.S. or 
opacity calculations. Molecules are fully dissociated in stars earlier than F0, 
which \citet{cox00} lists as $T_{\mathrm{eff}} \,\, \gtrsim \,\,$7300 K; we choose to 
err on the side of including molecules unnecessarily and include them in models 
with $T_{\mathrm{eff}} \,\, \leq \,\,$7500 K, to ensure they are present in all 
models where they are significant. We also take this as the $T_{\mathrm{eff}}$ 
value above which the possibility of convection is no longer considered in our models, 
treating the whole extent of the atmosphere as if it were in radiative equilibrium.
The one exception to this division of warm and cool stars is modeling the linear 
Stark broadening of H I lines. We include the broadening in the spectra of atmospheres 
with $T_{\mathrm{eff}} \,\, \geq \,\,$5000 K; spectra cooler than this do not 
have significant Stark wings on H I features.  

\paragraph{}

All of our atmospheric models have alpha enhanced abundances, with $\alpha$ = 
+0.4 dex. For our initial solar composition, we take our abundances up to O from 
\citet{grevesse96}, and take the revised abundances of \citet{scott15a} (F to Ca), 
\citet{scott15b} (Sc to Ni), and \citet{grevesse15} (Cu to Cs).  
We assume values for microturbulent velocities of $\xi \,\, = \,\, 4 \,\,\mathrm{km\,s^{-1}}$ 
for stars of log $g$ = 3.0 and lower, and $\xi \,\, = \,\, 1 \,\,\mathrm{km\,s^{-1}}$ 
for stars of log $g$ = 3.5 and higher. Two distinct values are used here instead of a 
more realistic continuous variation to artificially enhance the distinction between 
spectral lines dominated by evolved and unevolved populations in the IL spectrum. 

\paragraph{}

The synthetic spectral output of PHOENIX is the monochromatic flux spectral energy 
distribution (SED) of a model atmosphere, measured at an optical depth surface of 
$\tau_{12000}$ = 0. Each synthetic spectrum in our library needs to be scaled by 
a factor of $(R_{\tau=0}/R_{\tau=1})^2$, where the radii are obtained from the 
corresponding structural models, to convert to the flux spectrum at the $\tau_{12000}$
= 1 surface. This provides consistency between models of equivalent $T_{\mathrm{eff}}$ 
that have varying values of log $g$. We sample our spectra over the wavelength range 
$\lambda$ = 2000 to 27000 \AA~ at a spectral resolution of R = 300000, allowing 
us to compare values of cluster parameters derived from UV \citep{bellini15,piotto15} 
and IR \citep{cohen15,valcheva15} photometry, to those derived from more traditional 
optical photometry and spectroscopy.  

%(Still need the optical photometry and spectroscopy references here)

\subsection{NLTE Atmospheres \label{nlte}}

In this work, we explore the effects of NLTE atmospheric modeling on synthetic 
IL spectra and cluster ages and metallicities derived from them. The atmospheric 
structures and synthetic spectra are self-consistently modeled in NLTE. We focus 
on NLTE modeling because Fe I line strengths and EWs in an IL spectrum are the 
main diagnostic features for deriving cluster parameters \citep{mcwilliam08}, and 
Fe I is one of the atomic species most heavily affected by NLTE in synthetic spectra.

\subsubsection{NLTE Model Atoms \label{atoms}}

Because we study NLTE effects on GC parameters derived from IL spectra, the accuracy 
and completeness of our NLTE treatment is an important concern. It was shown by 
\citet{mashonkina11} that using a more complete Fe I atomic model will reduce NLTE
overionization effects by providing more high energy excited states to facilitate 
recombination from Fe II. They found that the greater the number of energy 
levels within $\Delta E \; = \; k T_\mathrm{eff}$ of the ground state ionization 
energy $(\chi_\mathrm{Ion})$, the more accurate the NLTE ionization equilibrium 
solution. 

\paragraph{}

To this end, we have adopted a set of new and updated NLTE model atoms, that are improved with respect to \citet{young14}.
These new atoms generally add to the numbers and refine the exact atomic data values of the energy 
levels and transitions over the old model atoms. Table \ref{tab:nlte species list} 
shows a comparison of the numbers of levels and lines between the old and new model atoms. In addition to the species 
listed in the table, H I, He I, and Ne I are also treated in NLTE, but are handled 
internally by PHOENIX and have not been updated. However, in the cases of He I and Ne I, 
these species do not dominate massive line blanketing, while the old H I treatment 
is sufficiently complete to model the asymptotic convergence of the Balmer 
series lines for rounding out the Balmer jump. 
For details on the NLTE treatment of these species, see \citet{young14}.

\paragraph{}

These new model atoms provide a significant improvement for some species, such as 
Fe I and II. Specifically, for Fe I, the new model atom has nearly double the 
number of energy levels, and more than triple the number of b-b transitions than 
the old. Additionally, the difference between the highest energy level and 
$\chi_\mathrm{ion}$ in the old Fe I model atom was $\Delta E \; = \; 0.322 eV$, 
which meant that stars cooler than $T_\mathrm{eff}\;\sim\;3750\,K$ would not have 
any energy levels within $kT_\mathrm{eff}$ of $\chi_\mathrm{ion}$. 
With the new model atom, there are now 45 energy levels closer to $\chi_\mathrm{ion}$, 
and for the coolest stars in our library at $T_\mathrm{eff}\;=\;3000\,K$, 17 energy 
levels are within $kT_\mathrm{eff}$ of $\chi_\mathrm{ion}$.

\section{SYNTHETIC COLOR MAGNITUDE DIAGRAMS \label{cmd}}

% short introductory paragraph about synthetic CMD, not observed

\subsection{Isochrones}

For this work, we have employed the Teramo theoretical isochrones, from the BaSTI 
group \citep{pietrinferni06}. The Teramo isochrones are offered with a variety of 
theoretical assumptions made in calculating the stellar evolutionary tracks, 
including alpha-enhanced or scaled solar compositions, with or without convective 
core overshooting, two different mass-loss rates following the Reimers Law, and 
normal or extended AGBs. The isochrones cover ranges of -3.27 to 0.51 dex in metallicity
and 30 Myr to 19 Gyr in age. Each one is sampled at 2000 mass points with a variable sampling frequency
to ensure that each area of the CMD is critically sampled. The 30 Myr isochrones 
cover a range of initial masses from 0.5 to 8.5 M$_\odot$, with the upper mass limit being reduced for isochrones with 
greater ages as stars evolve beyond the modeled tip of the AGB. The mass 
sampling frequency is increased for these isochrones to maintain the 2000 sampling points.

\paragraph{}

We select a subset of isochrones to investigate, with alpha enhancement of 
$\alpha \, = \, +0.4$, mass loss parameter $\eta \, = \, 0.2$, normal AGBs, and 
without core overshoot, similar to the selection of \citet{colucci09}. We focus our investigation on those isochrones with ages 
and metallicities covering the range of observed Galactic GC values, spanning 
9 to 15 Gyr and [M/H] = -1.49 to -0.66 dex.  These values were chosen as the 
isochrone metallicities closest to the peak values of the metal-rich and metal-poor 
Galactic GC populations \citep{zinn85}. We extend our ages beyond the range 
of average Galactic values (10-13 Gyr), to investigate the possible size of the 
effect NLTE modeling can have on derived ages.

\paragraph{}

The isochrones include stars mapping out the transition from the tip of the RGB 
to the red end of the HB. We remove these stars before populating our CMD 
for two reasons: 1) This transition is not well understood theoretically; and 2) 
Including these stars would interfere with 
our CMD discretization procedure, outlined below in Section \ref{boxing}. 
When included in a population, these stars contribute $< \, 1 \, \%$ of the total 
cluster luminosity, and have a negligible impact on the IL SED.

\subsection{Initial Mass Function}

To expand the isochrones into full populations, we use Kroupa's initial mass 
function (IMF) \citep{kroupa01}, normalized as a probability density function of the form
\begin{equation}
  p_\mathrm{Kroupa}(m)=\begin{cases}
    Ak_0m^{-0.3} & 0.01 \, \mathrm{M}_\odot < m < m_1\\
    Ak_1m^{-1.3} & m_1 < m < m_2\\
    Ak_2m^{-2.3} & m_2 < m < m_3\\
    Ak_3m^{-2.3} & m_3 < m
  \end{cases}
\end{equation}
with $k_0 \, = \, 1$, $k_1 \, = \, k_0m^{-0.3+1.3}_1$, $k_2 \, = \, k_1m^{-1.3+2.3}_2$,
and $k_3 \, = \, k_2m^{-2.3+2.3}_3$ ensuring a continuous function, where 
$m_1 \, = \, 0.08 \, \mathrm{M}_\odot$, $m_2 \, = \, 0.50 \, \mathrm{M}_\odot$,
and $m_3 \, = \, 1.00 \, \mathrm{M}_\odot$ \citep{maschberger12}. $A$ is a global 
normalization constant. This form gives information about the relative frequencies 
of stars of various masses as opposed to the number of stars of different masses in 
a unit spatial volume. 

\paragraph{}

The normalization constant, $A$, is determined for each isochrone individually, 
relative to their respective mass sampling ranges. To get the relative frequencies 
of the stars in each isochrone independently, we take a continuous mass range and 
divide it into bins centered on the isochrone points, with bin divisions halfway 
between adjacent points. The IMF is then numerically integrated over these 
bins using the extended trapezoid rule to get the relative frequency for each.

\subsection{Populating the CMDs}

To build our synthetic populations, we take a given target luminosity for the 
population, $L_\mathrm{tot}$, representative of the luminosity of a real cluster,
and analytically allocate fractions of $L_\mathrm{tot}$ 
to the isochrone points according to the relative frequencies. This determines the 
total luminosity of each point and, when divided by a point's individual luminosity, 
the number of stars representing that point in the population. This can result in 
non-whole numbers of stars for each isochrone point. 

\paragraph{}

There is no intrinsic spread of CMD features or observational scatter from any 
source inherent in the synthetic population of our CMDs; they are built as simple stellar 
populations out of single isochrones. We choose not to introduce any spread 
or scatter in the population artificially. This is unnecessary, as any random variations 
introduced will be averaged away in the creation of the IL spectrum, 
outlined in Section \ref{boxstars}.

\section{SYNTHETIC INTEGRATED LIGHT SPECTRA \label{ilspec}}

The integrated light spectrum of a globular cluster is the combined light of every individual star
within the cluster. Since it is not feasible to model hundreds of thousands of stars 
for each cluster that is to be studied, even when interpolating within a library
of stellar models, a method by which an IL spectrum can be approximated is necessary.

\subsection{Discretizing the CMD \label{boxing}}

We choose to represent groups of parametrically similar stars in a GC CMD by 
a single stellar spectrum per group and weight their contributions to the IL 
spectrum, following the method of \citet{mcwilliam08}. The method involves
discretizing a CMD by binning parametrically similar stars areas 
bounded by lines of constant $T_{\mathrm{eff}}$ and $L_\mathrm{Bol}$. These areas, 
or boxes, are established such that an approximately equal fraction of the total 
cluster luminosity is emitted by the stars contained in each.  \citet{mcwilliam08} 
limit the luminosity contained in any box to $\sim$ 3 to 4 \% of the total luminosity 
for a given cluster ($\sim$ 25 to 30 boxes total); we choose to increase the total 
number of boxes to 50 ($\sim$ 2 \% of $L_\mathrm{tot}$ each), effectively doubling 
the discretization resolution in our CMDs. This prevents the boxes from covering 
too large a range of values of the stellar parameters, which ensures that a single 
representative stellar spectrum will be an accurate approximation of the integrated 
spectrum for a box. 

\paragraph{}

\begin{figure}
\plotone{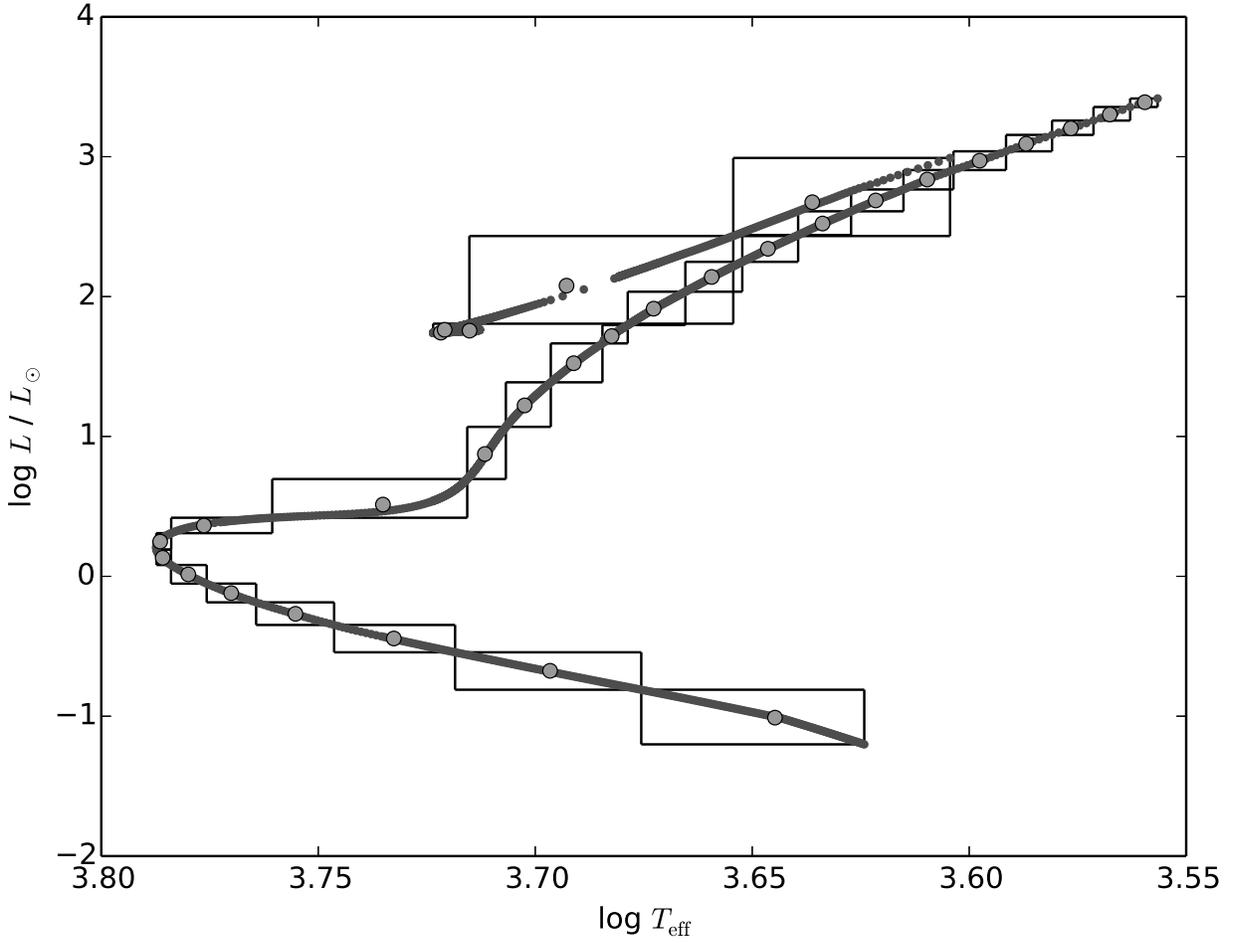}
\caption{CMD of a sample population similar to 47 Tuc (age = 13.0 Gyr, [M/H] = -0.66). The black outlines are our CMD discretization ``boxes'', the dark gray points are isochrone sampling points, and the light gray circles are box representative average stars.\label{fig:samplecmd}}
\end{figure}

Starting at the low-mass end of the main sequence, stars of increasing mass are 
added to a box until the sum of their luminosities matches the allotted percentage 
of the total cluster luminosity for a box. The box is then considered full and 
subsequent stars are added to a new box. The process is repeated for increasing 
stellar mass until everything up to the tip of the RGB is enclosed in a box. 
We repeat the process starting at the low-mass end of the HB and proceed up the 
AGB, until the entire population has been placed into boxes. The process is 
explicitly broken and restarted here to avoid too large of a box from encompassing 
both the RGB tip and HB by blindly adding stars of increasing masses to boxes.
This effectively splits our CMD into two components by ignoring short lived and 
poorly understood stars in the transition from the RGB to the HB. The first 
component is comprised of the main sequence, sub-giant branch, and red giant branch
(the MR component), while the second encompasses the asymptotic giant and horizontal 
branches (the HA component). Figure \ref{fig:samplecmd} shows an example of a 
theoretical population similar in age and metallicity to 47 Tuc (age = 13.06 Gyr,
[M/H] = -0.78) \citep{forbes10}, including the discretization boxes. 

\paragraph{}

To determine the number of boxes necessary to properly sample the CMD, while 
maintaining approximately equal luminosity for each box, we calculate the ratio 
of the luminosity of the MR component ($L_\mathrm{MR}$) to that of the HA component 
($L_\mathrm{HA}$). We then iterate through a total number of boxes, as well as 
the numbers of boxes allotted to each component, finding the optimal combination 
to be the one where the ratio of MR boxes ($N_\mathrm{MR}$) to HA boxes ($N_\mathrm{HA}$) 
most closely matches the ratio of luminosities. By limiting the total number of 
boxes to between 25 and 50, this combination gives the closest agreement between 
$$\frac{L_\mathrm{MR}}{N_\mathrm{MR}} \; \sim \; \frac{L_\mathrm{HA}}{N_\mathrm{HA}}$$
maintaining approximately equal luminosity in each box, while still limiting each 
box to 2 to 4 \% of the total cluster luminosity.

\subsection{Creating the Representative Stars \label{boxstars}}

Once the CMD discretization is completed, we proceed to calculate the atmospheric modeling 
parameters for a representative star in each box. To produce a synthetic spectrum for this 
box star, we require $T_{\mathrm{eff}}$, log $g$, [M/H], and either $M$ or $R$.
Metallicity is constant along each isochrone, and prescribes the value for a box
star directly. As we are interested in representing the combined light of every 
star in a box with a single stellar spectrum, we take the average of 
the parameters over a whole box, weighted by 
$$w_* \; = \; N_* \; / \; N_{\mathrm{box}}$$
the fractional number of stars of each isochrone mass sampling bin per box. 

\paragraph{}

The $T_{\mathrm{eff}}$ of our box star is found from 
$$<T_{\mathrm{eff}}> \; = \; <w_* \; * \; T_{\mathrm{eff,*}}^{\;\;\;\;4}>^{1/4}$$
where the values for individual stars, $T_{\mathrm{eff,*}}$, are taken directly
from the isochrones. We take the average of the fourth power, rather than a linear 
average, to include the relative contribution each star makes to the total 
luminosity of a box. Similarly, the average mass and luminosity can be found from 
$$<M> \; = \; <w_* \; * \; M_*>$$
$$<L_{\mathrm{bol}}> \; = \; <w_* \; * \; L_{\mathrm{bol},*}>$$
where once again, the individual quantities are taken directly from the isochrones.

\paragraph{}

There is some ambiguity in calculating the average log $g$ for a box. Because log $g$
is not a stellar interior modeling parameter, it is not included directly in the
isochrones, so a direct average is unavailable in this case. We choose to take the 
averages of the relevant isochrone quantities and calculate a single $g$ from those average values,
$$<g> \; = \; \frac{G \; <M> \; 4\pi \; \sigma \; <T_{\mathrm{eff}}^{\;\,4}>}{<L>}$$
without calculating individual $g$ values for the isochrone sampling points.
This method has the benefit of being consistent with the reverse process of observers inferring parameters  
from the IL spectrum of a group of spatially unresolved stars.

\subsection{Synthesizing IL Spectra}

\begin{figure}
\plotone{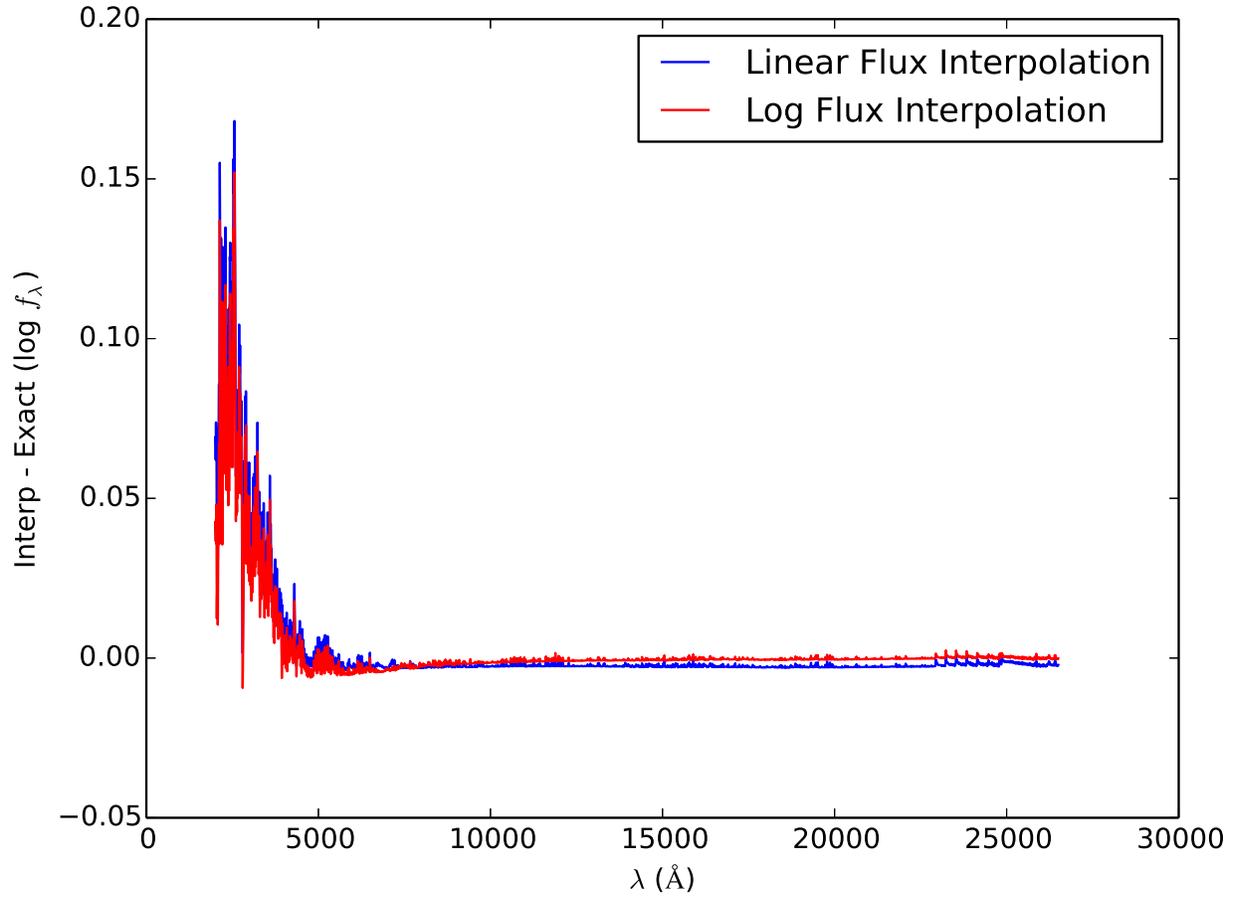}
\caption{The differences between IL spectra generated by interpolating library spectra and an IL spectrum generated from spectra with the exact parameters of the representative stars. $Blue$ - Linear interpolation of linear fluxes. $Red$ - Linear interpolation of log fluxes.\label{fig:interpolation}}
\end{figure}

Now that the full CMD distribution has been reduced to the representative stars, we generate 
stellar spectra for each box by interpolating among synthetic spectra in our 
library. We linearly interpolate our library spectra
weighted by three or four atmospheric modeling parameters; $T_{\mathrm{eff}}$, 
$\mathrm{log}\,g$, M, and, in cases where the isochrone value is not a direct 
match to one of the values in our library, [M/H]. This interpolation scheme results 
in each box spectrum being formed by interpolating among either 8 (matching 
library [M/H]) or 16 (interpolating [M/H]) individual spectra from our 
library. We chose to interpolate using a linear method because our library 
is already pushing the lower boundaries of atmospheric structure convergence in 
$T_{\mathrm{eff}}$ and $\mathrm{log}\,g$, and higher order methods would require 
additional synthetic spectra with even lower values for these parameters. We test 
the accuracy of this interpolation by comparing two IL spectra generated using 
this procedure (one interpolating linear flux spectra and one interpolating 
log flux spectra) to one generated from synthetic spectra with the exact parameters of 
the representative stars. Figure \ref{fig:interpolation} 
shows that there is relatively little difference between our interpolated and exact 
IL spectra, except for the shortest wavelengths that we model, and that there is 
little appreciable difference between interpolating linear or log fluxes.
Representative populations for the youngest and oldest isochrones in 
this study are plotted in Figure \ref{fig:library} as a visual indication of where the 
interpolation between library spectra will occur for $T_\mathrm{eff}$ and $\mathrm{log}\,g$.

\paragraph{}

In nature, IL spectra are combinations of the luminosity spectra of individual 
stars, not the flux spectra. Because of this, we must convert our box representative 
spectra from PHOENIX surface fluxes to luminosities. The most direct method of 
doing so takes advantage of 
$$\frac{L_\lambda}{F_\lambda} \; = \; \frac{L_{\mathrm{bol}}}{F_{\mathrm{bol}}}$$
where $L_{\mathrm{bol}}$ for a box is the average value as calculated above, and we 
calculate the $F_{\mathrm{bol}}$ by numerically integrating low resolution PHOENIX 
spectra from 10 to 10,000,000 \AA~ using the extended trapezoid rule, ensuring sufficient 
coverage of both the Wien side and Rayleigh-Jeans tail of the spectra. 

\paragraph{}

\begin{figure}
\includegraphics[scale=.75]{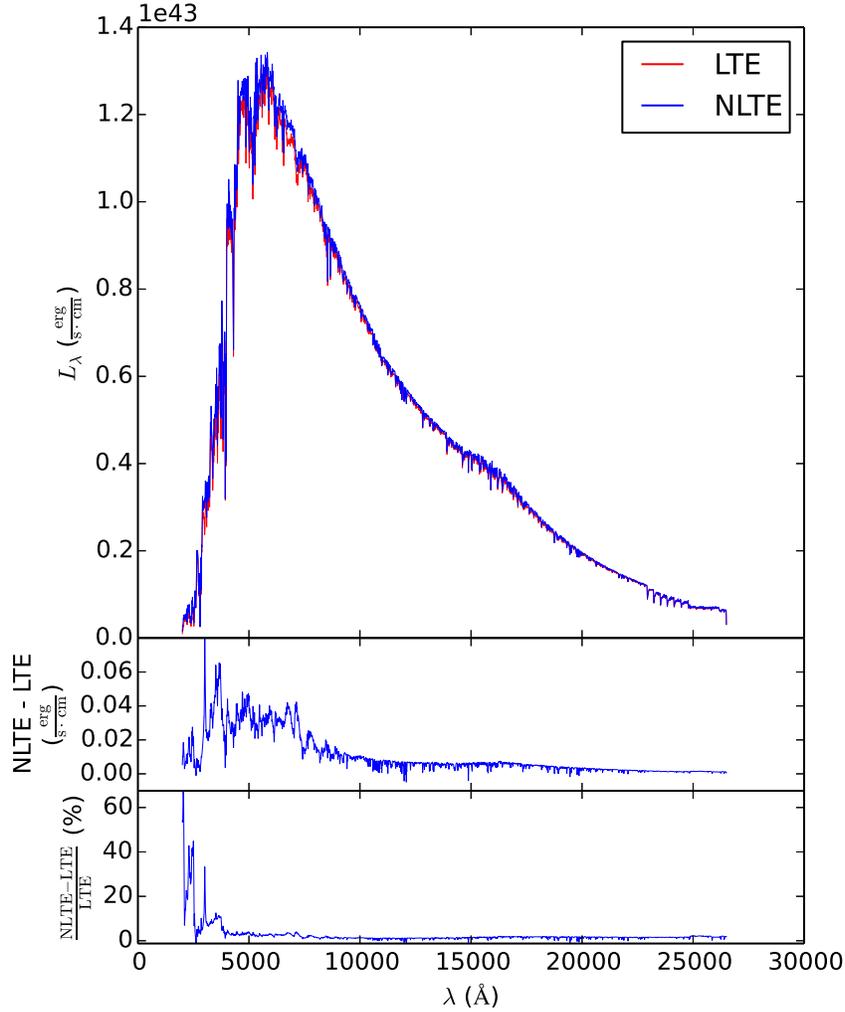}
\caption{Comparison of NLTE and LTE IL spectra for the population of Figure \ref{fig:samplecmd}. The species responsible for the majority of discrepancies between the spectra are light metals, primarily Fe I, with a few exceptions. The large deviation observed in the range from $\lambda\,\approx$ = 4000 to 8500 \AA~ is caused by TiO molecular bands. The clusters of lines seen in the range $\lambda\,\approx$ = 10000 to 12000 \AA~ are those of Ti I, and the strong lines near $\lambda\,\approx$ = 19000 to 20000 \AA~ are those of Ca I. All three panels have been convolved from our high resolution spectra to a spectral resolution of R $\sim$ 5000 for ease of viewing. $Top$ - Synthetic IL spectra for the population of Figure \ref{fig:samplecmd}. $Middle$ - Absolute difference between NLTE and LTE synthetic IL spectra. The NLTE spectrum is more luminous in the UV than LTE, while showing stronger absorption features in the IR. $Bottom$ - Relative difference between NLTE and LTE synthetic IL spectra. \label{fig:ilspec}}
\end{figure}

We now combine the box spectra into a synthetic IL spectrum. Each spectrum is 
scaled by $$ L_{\lambda,box} \; = \; N_{\mathrm{box}} \; * \; L_{\lambda}$$ to 
account for the total luminosity of the box, and then added together. Three IL spectra 
per CMD are created this way; one composed of LTE stellar spectra, one of NLTE 
stellar spectra, and one composed of both LTE and NLTE stellar spectra where only 
the evolved population were NLTE. For our purposes, we 
consider any star more evolved than the sub-giant branch to be ``evolved''. These 
hybrid IL spectra allow us to isolate and study the impact NLTE modeling of the 
evolved population has on cluster parameters inferred from IL spectra. The LTE 
and NLTE spectra for our 47 Tuc population are overplotted 
for comparison in Figure \ref{fig:ilspec}, with the absolute and relative NLTE-LTE 
differences, to highlight the most disparate spectral features.

\paragraph{}

\begin{figure}
\includegraphics[scale=.90]{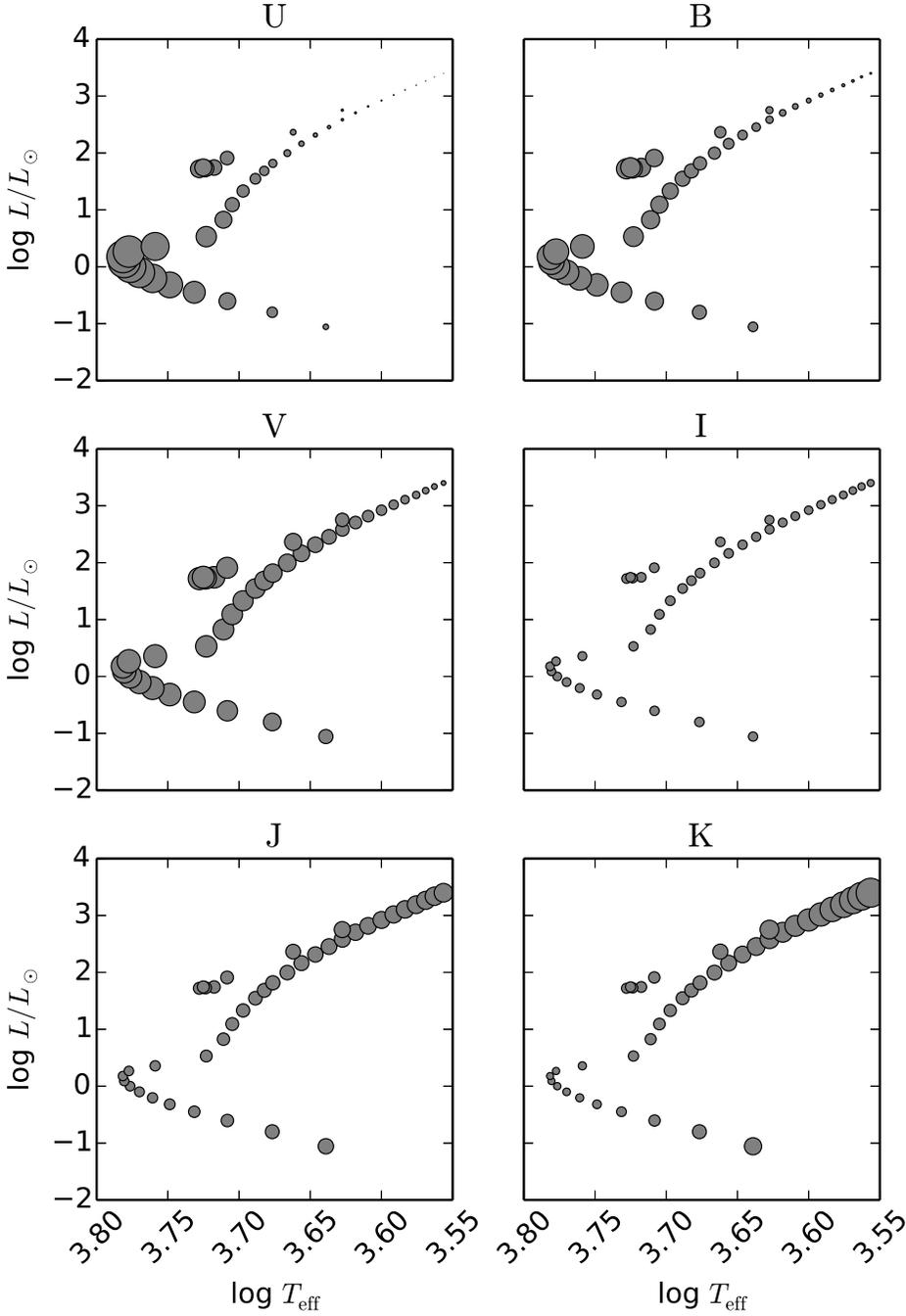}
\caption{CMDs of the 15.0 Gyr population representative stars, displaying each star's relative contribution to the IL spectrum in select photometric bandpasses. For each filter, the radius of each circle is scaled to the percentage of the total IL flux in that band contributed by the star. Circle sizes are not correlated across different filters. \label{fig:bubbles}}
\end{figure}

Identifying the atomic and molecular species responsible for the large
differences between the LTE and NLTE spectra is not a straightforward task when 
dealing with IL spectra. Because an IL spectrum is the co-added light of multiple 
spectra of different spectral types, what appears to be a single feature in the 
IL spectrum may be caused by multiple sources of opacity. Additionally, at the 
modest spectral resolution of Figure \ref{fig:ilspec}, blending of features in 
crowded regions can confuse the issue even further.
To proceed with IL feature identification, we take five sample stars from our 
library models ($T_\mathrm{eff}$ = 6500 K and $\mathrm{log}\,g$ = 4.5, $T_\mathrm{eff}$ 
= 5750 K and $\mathrm{log}\,g$ = 3.5, $T_\mathrm{eff}$ = 5000 K and $\mathrm{log}\,g$ 
= 3.0, $T_\mathrm{eff}$ = 4250 K and $\mathrm{log}\,g$ = 2.0, $T_\mathrm{eff}$ = 
3600 K and $\mathrm{log}\,g$ = 0.5), and identify the sources of any large 
discrepancies between the sample NLTE and LTE spectra. 
We weight a NLTE-LTE difference in the spectrum of a sample star by that star's 
relative contribution to the IL luminosity in the photometric band corresponding to 
the wavelength of that difference.
Figure \ref{fig:bubbles} displays 
an example of these relative contributions for the 15.0 Gyr population. 
If the discrepancies in the sample stars are also found to be present in the IL 
spectrum, the sources in the 
individual spectra are considered to be responsible. The species responsible for 
the majority of large differences were found to be primarily light metals in their 
ground states, Fe I being the most prominent among these, with a few exceptions. 
The large deviation observed in the range from $\lambda\,\approx$ = 4000 to 8500 
\AA~ was identified as TiO molecular bands from our cool giant stars. The clusters 
of lines seen in the range $\lambda\,\approx$ = 10000 to 12000 \AA~ were Ti I, 
and the strong lines near $\lambda\,\approx$ = 19000 to 20000 \AA~ were 
found to be Ca I. A full set of high resolution diagnostics will be included in 
a forthcoming paper in this series.

\section{PHOTOMETRIC ANALYSIS \& RESULTS \label{photo}}

To take an initial estimate of the magnitude of the effect NLTE modeling has on 
parameters derived from IL spectra, we examine four photometric color indices 
($\mathrm{U-B}$, $\mathrm{B-V}$, $\mathrm{V-I}$, and $\mathrm{J-K}$) produced using 
Bessel's updated Johnson-Cousins UBVRI photometric system \citep{bessel90} as well 
as Bessel and Brett's VJHKLL'M photometric system \citep{bessel88}. We analyze a 
collection of IL spectra for synthetic CMDs approximating 47 Tuc, with [M/H] 
= -0.66 and substituting $M_v$ = -8.64 for $L_{\mathrm{tot}}$, except ranging in 
age from 9 to 15 Gyr. Two sets of these spectra are generated, each using the LTE, NLTE, 
and hybrid prescriptions, limiting the number of boxes to 25 to 35 (set 1) and 40 to 50 (set 2). 
Synthetic photometric colors for these IL spectra are single-point
calibrated to both a NLTE synthetic spectrum approximating Vega ($T_{\mathrm{eff}}$ = 
9600 K, $\mathrm{log}\,g$ = 4.1, [M/H] = -0.5), and the library spectrum 
that most closely approximates Arcturus ($T_{\mathrm{eff}}$ = 4250 K, 
$\mathrm{log}\,g$ = 2.0, [M/H] = -0.66). Arcturus was chosen for this second 
calibration to compare the IL colors to that of a standard star that is representative 
of the populations. We denote color values calibrated to each star as 
$\mathrm{X-Y}_\mathrm{Vega}$ and $\mathrm{X-Y}_\mathrm{Arc}$ respectively.

\subsection{Colors and Ages \label{colors}}

\begin{figure}
\includegraphics[scale=.45]{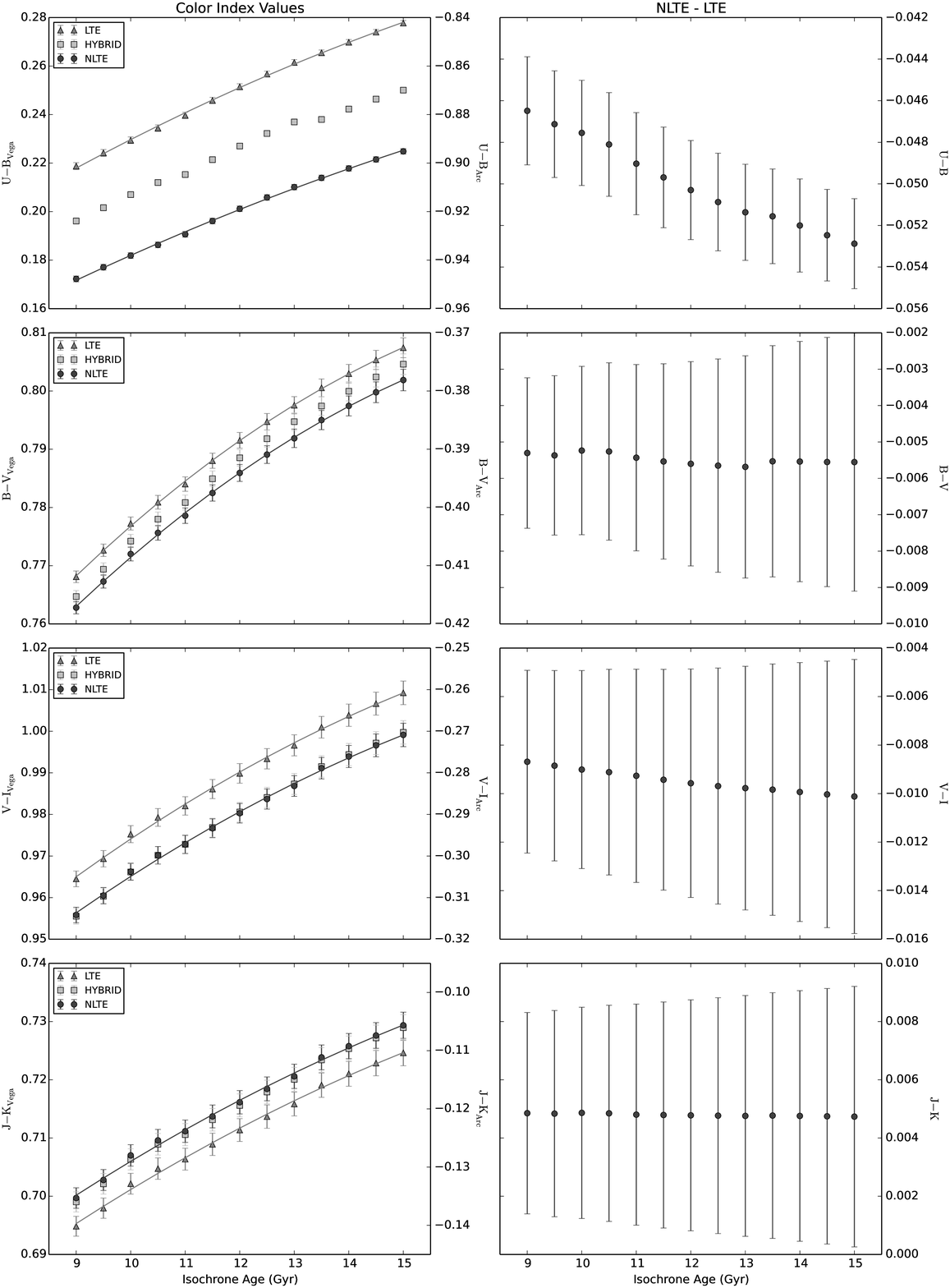}
\caption{$Left\,Column$ - Photometric colors for synthetic IL spectra of populations with constant bolometric luminosity, and [M/H] = -0.66. All populations are discretized with 25 to 35 boxes. The error bars are the CMD discretization uncertainty, outlined in Section \ref{uncertainty}. Solid lines are second order polynomials used for parameterizing the data.  $Right\,Column$ - Difference between NLTE and LTE photometric colors. \label{fig:boxes25}}
\end{figure}

\begin{figure}
\includegraphics[scale=.45]{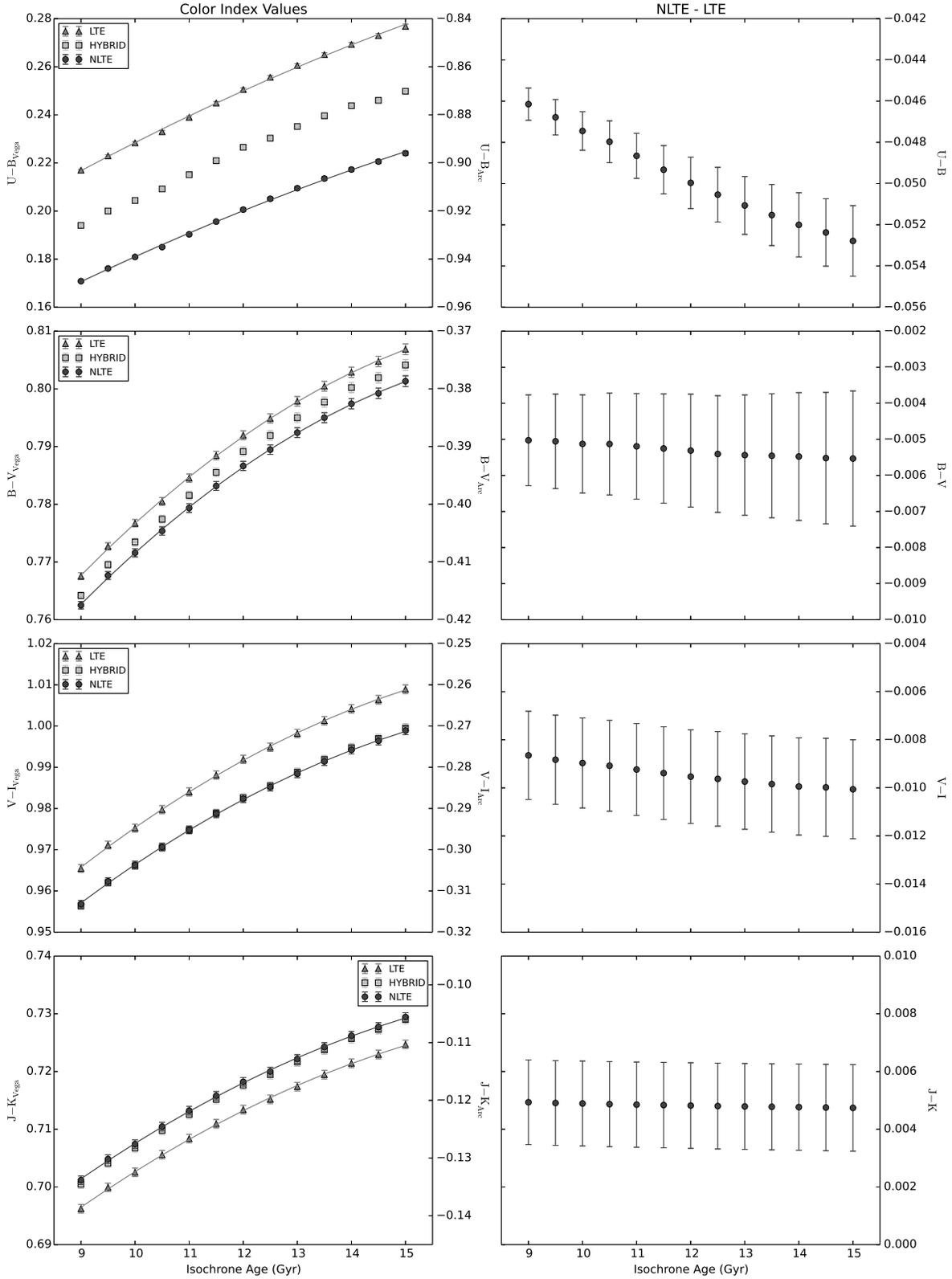}
\caption{Similar to Figure \ref{fig:boxes25}, but for populations discretized with 40 to 50 boxes. \label{fig:boxes50}}
\end{figure}

Figures \ref{fig:boxes25} and \ref{fig:boxes50} present the values of the 
color indices as a function of population age for sets 1 and 2 respectively, 
highlighting the NLTE - LTE differences. Qualitatively, there is little difference
between the two sets. All four color indices display a reddening of IL color as 
the population ages, as more of the population evolves into older, cool stars. 
The NLTE colors for the first three color indices are bluer than LTE at 
all ages, because NLTE overionization of Fe I weakens the myriad weak Fe I 
lines that have the character of a pseudo-continuous opacity in the blue and near 
UV bands. Conversely, the $\mathrm{J-K}$ NLTE colors are redder than LTE, where 
the surplus of free electrons produced by the overionization increases H$^-$ opacity 
in the J band ($<$ 1.6 $\mathrm{\mu m}$). The NLTE-LTE color differences for the 
indices in both sets are consistent with a constant value as a function of age, 
with the exception of $\mathrm{U-B}$, where the difference is seen to increase as 
a function of age.

\paragraph{}

The hybrid color values fall midway between the LTE and NLTE colors for the UV 
and optical indices, and converge with the NLTE values at IR wavelengths. 
There are two possible explanations for this convergence: 1) Evolved stars dominate 
the IL spectra in these filters' wavebands; and 2) NLTE effects in main sequence 
stars negligible in these wavebands. To determine which of the two effects is 
responsible for the convergence, we refer back to the representative stars' relative 
luminosity contributions in Figure \ref{fig:bubbles}. As can be seen for 
$\mathrm{V-I}$, evolved stars do not dominate the IL spectrum, with all 
representative stars making approximately equal contributions to the IL spectrum 
in the $\mathrm{I}$ band, and the main sequence stars near the turnoff 
are the strongest contributors in the $\mathrm{V}$ band. In this case, it would 
suggest that negligible NLTE effects in main sequence stars at these wavelengths 
are responsible for the observed convergence of hybrid to NLTE colors. For 
$\mathrm{J-K}$, the same explanation is likely responsible, but a combination of 
the two effects is also possible. For both the $\mathrm{J}$ and $\mathrm{K}$ bands, 
evolved stars on the red giant branch are the strongest contributors to the IL 
spectrum (although they do not dominate it).

\paragraph{}

To give a quantitative estimate for how much of an impact NLTE effects in IL 
spectra have on derived ages, we first define the quantity $\Delta_\mathrm{Age}$, 
the uncertainty in derived age from the uncertainty in measured color index value, such that
$$\Delta_\mathrm{Age} \; = \; \Delta_\mathrm{color} \frac{\mathrm{d} A(\mathrm{color})}{\mathrm{d(color)}}$$
where $\Delta_\mathrm{color}$ is the numerical uncertainty of a given color value, 
and $A(\mathrm{color})$ describes the derived cluster age as a function of ``observed'' color 
value, found by parameterizing the IL color vs age relation with a low order polynomial. We find 
that a linear function does not provide a good match to the relation, but a parabola 
provides an excellent match for all color indices, with coefficients of determinations 
of at least $R^2\geq0.997$. We contrast these $\Delta_\mathrm{Age}$ values with the difference 
between the NLTE and LTE derived ages for a given color, $A_\mathrm{color,NLTE}-
A_\mathrm{color,LTE}$. The full list of $\Delta_\mathrm{Age}$ values and the 
differences in NLTE and LTE derived ages for all color indices are presented in 
Table \ref{tab:age errors}. For comparison, we also present the uncertainty 
in derived ages found from setting $\Delta_\mathrm{color}$ equal to 0.01 mag, the limiting
precision of the Harris catalog \citep{harris96}.

\paragraph{}

The $\mathrm{U-B}$ color index returned the largest differences, but these results 
were considered to be unrealistic for a number of reasons, including keeping the 
metallicity fixed throughout this experiment and the difficulty associated with 
modeling the near-UV region of stellar spectra.
The other three indices, $\mathrm{B-V}$, $\mathrm{V-I}$, and $\mathrm{J-K}$, 
produced comparable age differences, with those derived from $\mathrm{V-I}$ 
generally being the largest by $\lesssim$ 40 \%. 
Age differences for our $\mathrm{B-V}$ index ranged from 0.61 to 1.58 Gyr for set 
1, and 0.55 to 2.53 Gyr for set 2. The smaller age difference for each set corresponds 
to the bluest color measured for the LTE IL spectra, and the larger difference 
corresponds to the reddest measured LTE color. For comparison, the range of 
$\Delta_\mathrm{Age}$ values for $\mathrm{B-V}$ are 0.22 to 0.82 Gyr and 0.13 to 0.53 
Gyr, for sets 1 and 2 respectively. Assuming an observational limiting precision 
of 0.01 mag for $\mathrm{B-V}$, the limiting precision of the Harris catalog,  
results in $\Delta_\mathrm{Age}$ values of 1.12 to 2.41 Gyr and 1.05 to 2.84 Gyr.

\paragraph{}

We note that all of our IL spectra were produced from isochrones of fixed metallicity, 
and that age estimates derived in this fashion may vary greatly with changing metallicity.
It should also be noted that both LTE and NLTE model atmospheres have been shown to 
overpredict near-UV flux in the spectra of cool giant stars \citep{short09}, and this will likely 
be reflected in the IL spectra. Any results found from fitting synthetic IL 
spectra to observed spectra at UV wavelengths would be impacted by this overprediction.
Either or both of these effects may help explain the large differences in ages 
derived from our $\mathrm{U-B}$ colors. Additionally, this work is only 
concerned with a differential analysis between LTE and NLTE IL spectra, and in 
turn is only affected by any difference in the overpredictions. NLTE models are 
worse in the overprediction than their LTE counterparts as a result of NLTE Fe I 
overionization, but the updated Fe I NLTE model atom we have implemented should 
minimize this for the range of stellar parameters with which we are concerned. We also 
expect that, to first order, the NLTE-LTE difference in the overprediction of UV 
flux to be constant as a function of isochrone age.

\subsection{Numerical Uncertainty \label{uncertainty}}

\begin{figure}
\plotone{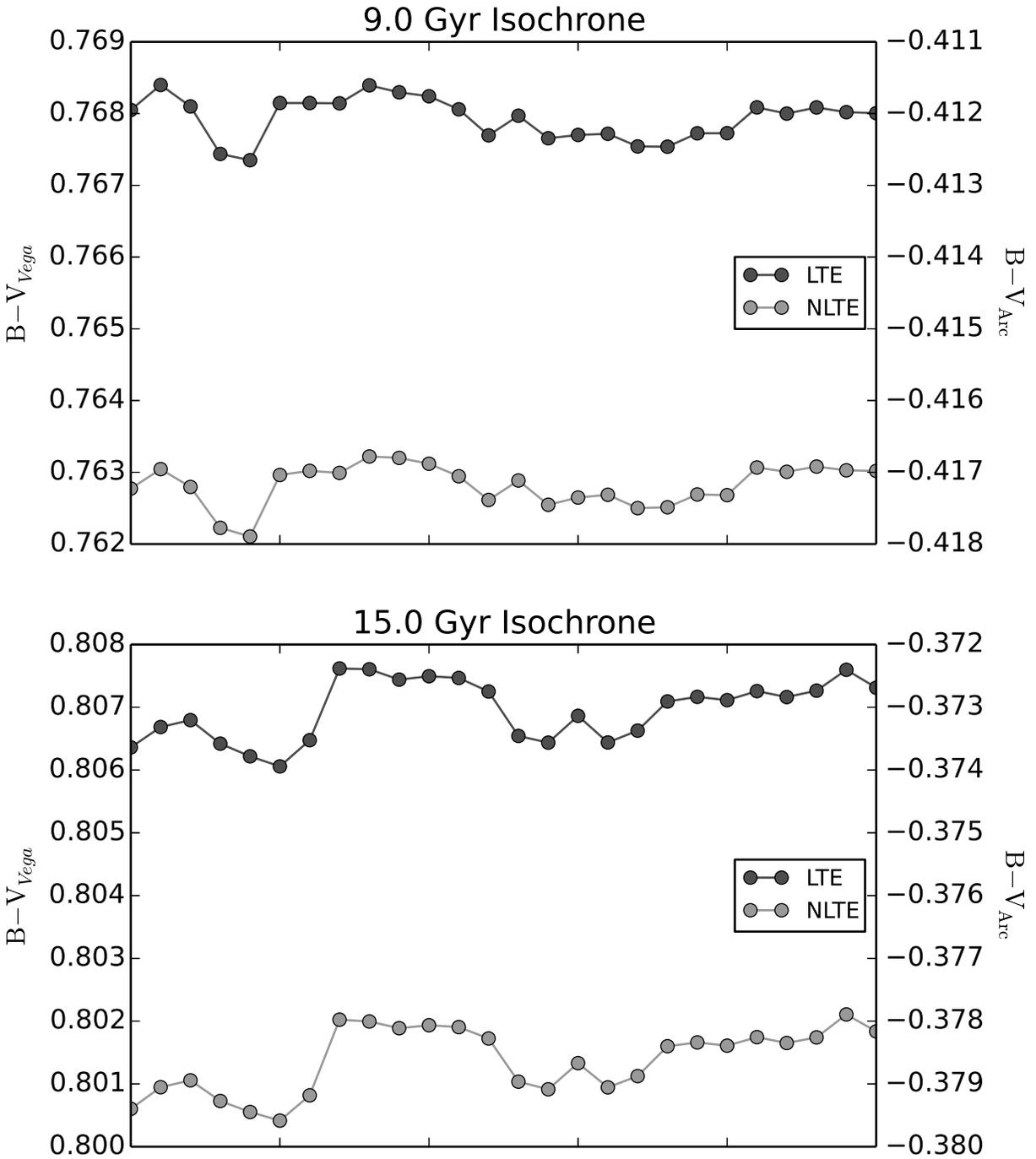}
\caption{Variation of $\mathrm{B-V}$ color with CMD discretization resolution. $Top$ - 9.0 Gyr population. $Bottom$ - 15.0 Gyr population. \label{fig:uncertainty}}
\end{figure}

Figure \ref{fig:uncertainty} shows the $\mathrm{B-V}$ colors for two isochrones, 
ages 9 and 15 Gyr, as a function of the number of boxes used to discretize the CMD. 
The value of a color index for a given population varies at the millimagnitude 
level as a function of the CMD discretization resolution. As might be expected, the variations 
are larger for lower discretization resolution, and are reduced for higher resolution.  
This trend is qualitatively repeated for the other color indices. 

\paragraph{}

\begin{figure}
\includegraphics[scale=.75]{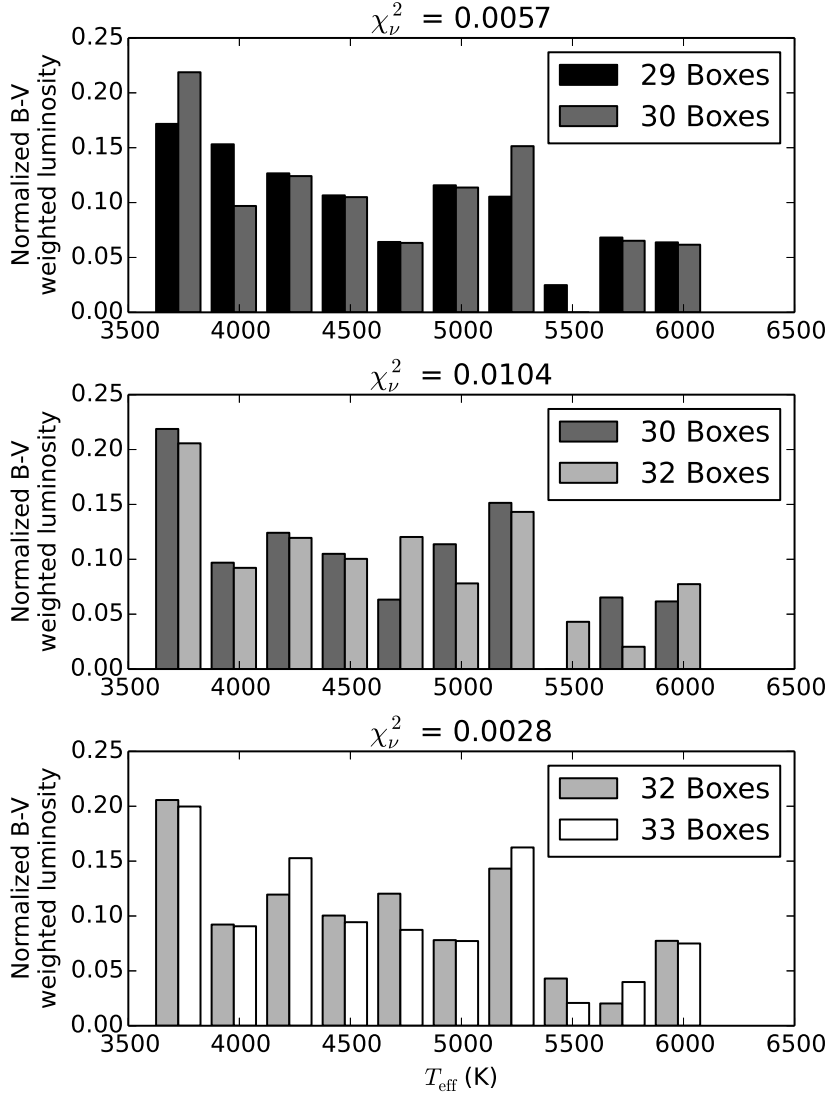}
\caption{Binned representative populations for the 15.0 Gyr isochrone, with different levels of CMD discretization. The bin size is set to 250 K, the $T_\mathrm{eff}$ resolution of our library. Each bin is weighted by a factor of $f_B/f_V$, the flux the bin contributes to the IL spectrum in the B band divided by the flux in the V band, representing the influence each bin has on the IL $\mathrm{B-V}$ value. The $\chi^2$ statistics, calculated in each case as $\Sigma\frac{(N-n)^2}{n^2}$, where N and n are the bin values for the sets with the greater and fewer number of boxes respectively, are used as a confirmation that the histograms in the middle panel exhibit greater differences than those in either the top or bottom panels. The middle panel, which compares discretization resolutions on either side of the large jump observed in Figure \ref{fig:uncertainty}, shows noticeable differences in bins with $T_\mathrm{eff}$ values $\geq$ 4750 K. \label{fig:histograms}}
\end{figure}

To isolate which regions of the CMD are being over- or under-sampled, 
we plot in Figure \ref{fig:histograms} the $T_\mathrm{eff}$ 
of the representative stars in three histograms, comparing the CMDs on either side 
of the largest change in color (ie. that between 30 and 32 boxes), and two control 
cases to either side of the largest change where there is relatively little change 
(29 to 30 boxes, and 32 to 33 boxes). The bin size for each histogram was set 
to 250 K, the temperature resolution of our library of spectra, and the bar heights 
for the four sets of representative stars (29, 30, 32, and 33 boxes) were each 
weighted by a factor of $f_B/f_V$, the flux in the B band emitted by stars in that
bin divided by the flux in the V band, the influence a given bin has 
on the IL $\mathrm{B-V}$ value, and independently normalized to sum to 1. 
We use a reduced $\chi^2_\nu$ statistic, listed with each panel in Figure 
\ref{fig:histograms}, to confirm that there is a greater difference between the 
populations in the 30 to 32 boxes histogram than in either the 29 to 30 boxes or 
the 32 to 33 boxes histograms. Inspection of this histogram reveals the most 
significant differences between the two populations occur for $T_\mathrm{eff}$ values $\geq$ 
4750 K. For our populations, this corresponds to the upper main sequence (including 
the turnoff), the base of the red giant branch, and the horizontal branch. Special 
care must be paid to these regions when discretizing the CMD to ensure they are 
not under-sampled.

\paragraph{}

As a measure of the numerical uncertainty in the IL spectrum resulting from CMD 
discretization, we evaluate the $3\sigma$ deviations in the computed integrated B-V 
index as a function of the number of boxes used to discretize isochrones at several ages spanning 
the age range. Uncertainties for the other color indices are obtained in a 
similar fashion. These uncertainties are represented as the error bars in Figures \ref{fig:boxes25} and \ref{fig:boxes50}.

\section{SUMMARY \label{summary}}

We have investigated a number of aspects of the method presented by \citet{colucci11}, 
refined from that of \citet{mcwilliam08}, for synthesizing GC populations and IL from a library 
of stellar atmospheric models and spectra. Following this method, a collection of 98 
IL spectra for clusters approximating 47 Tuc were generated with different CMD 
discretization resolutions and different degrees of NLTE treatment. 

\paragraph{}

For these clusters, age estimates that may be derived by fitting observed 
photometric colors with synthetic LTE colors were shown to differ from those 
similarly obtained from NLTE modeling by up to 2.54 Gyr. These 
age differences, while larger than the numerical uncertainties inherent in our 
methodology, are comparable with the limiting observational precision of current
catalogs.

\paragraph{}

Our investigation of CMD discretization resolution has shown that the IL spectrum is 
resolution-dependent, and that the effects on the spectrum are stronger at lower 
resolution. These effects are more prominent at shorter wavelengths. We also find 
that the 25 to 35 boxes recommended in the literature do not provide enough resolution 
to critically sample the upper main sequence and horizontal branch. At 
least 40-50 boxes are necessary to minimize the resolution dependency.

\paragraph{}

Initial analysis suggests that NLTE effects in MS stars have approximately 
equivalent influence on IL spectra as do those in evolved stars for UV and optical wavelengths, 
but negligible influence for IR wavelengths. This effect appears to be independent 
of CMD discretization resolution for those resolutions investigated here.

\subsection{Future Work}

Future papers in this series will be based on an expanded spectral library with 
a metallicity dimension that spans the range between the peaks of the bimodal 
distribution of Galactic GCs. We will expand the photometric
analysis to include the Hubble photometric system for additional UV colors, motivated 
by the fact that spectral features in the UV are highly sensitive to changes in metallicity. 
We will investigate equivalent widths and line profiles for nearly 600 spectral 
lines as well as Lick indices, all identified in the literature as useful features 
in determining cluster 
metallicities, for multiple atomic species, including Fe I and II. We also plan on 
expanding the methodology to include identifying multiple populations within single
clusters and deriving their parameters.

%% If you wish to include an acknowledgments section in your paper,
%% separate it off from the body of the text using the \acknowledgments
%% command.

%% Included in this acknowledgments section are examples of the
%% AASTeX hypertext markup commands. Use \url without the optional [HREF]
%% argument when you want to print the url directly in the text. Otherwise,
%% use either \url or \anchor, with the HREF as the first argument and the
%% text to be printed in the second.

\acknowledgments

We would like to thank the NSERC Discovery Grant program and Saint Mary's 
University's Faculty of Graduate Studies and Research for funding this work. 
We would also like to acknowledge Compute Canada member ACENet for providing us 
with all computational resources and CPU time.

\clearpage

%% Use the figure environment and \plotone or \plottwo to include
%% figures and captions in your electronic submission.
%% To embed the sample graphics in
%% the file, uncomment the \plotone, \plottwo, and
%% \includegraphics commands

%% If you need a layout that cannot be achieved with \plotone or
%% \plottwo, you can invoke the graphicx package directly with the
%% \includegraphics command or use \plotfiddle. For more information,
%% please see the tutorial on "Using Electronic Art with AASTeX" in the
%% documentation section at the AASTeX Web site,
%% http://www.journals.uchicago.edu/AAS/AASTeX.

%% The examples below also include sample markup for submission of
%% supplemental electronic materials. As always, be sure to check
%% the instructions to authors for the journal you are submitting to
%% for specific submissions guidelines as they vary from
%% journal to journal.

%% This example uses \plotone to include an EPS file scaled to
%% 80% of its natural size with \epsscale. Its caption
%% has been written to indicate that additional figure parts will be
%% available in the electronic journal.

\clearpage

\clearpage

%%%%%%%%%%%%%%%%%%%%%%%%%%%%%%%%%%%%%%%%%%%%%%%%%%%%%

\clearpage

\clearpage

%%%%%%%%%%%%%%%%%%%%%%%%%%%%%%%%%%%%%%%%%%%%%%%%%%%%%

\clearpage

\clearpage

%%%%%%%%%%%%%%%%%%%%%%%%%%%%%%%%%%%%%%%%%%%%%%%%%%%%%

\clearpage

\clearpage

%%%%%%%%%%%%%%%%%%%%%%%%%%%%%%%%%%%%%%%%%%%%%%%%%%%%%

\clearpage

\clearpage

%%%%%%%%%%%%%%%%%%%%%%%%%%%%%%%%%%%%%%%%%%%%%%%%%%%%%

\clearpage

\clearpage

%%%%%%%%%%%%%%%%%%%%%%%%%%%%%%%%%%%%%%%%%%%%%%%%%%%%%

\clearpage

\clearpage

%%%%%%%%%%%%%%%%%%%%%%%%%%%%%%%%%%%%%%%%%%%%%%%%%%%%%

\clearpage

\clearpage

%%%%%%%%%%%%%%%%%%%%%%%%%%%%%%%%%%%%%%%%%%%%%%%%%%%%%

\clearpage

\clearpage

%% Tables should be submitted one per page, so put a \clearpage before
%% each one.

%% Two options are available to the author for producing tables:  the
%% deluxetable environment provided by the AASTeX package or the LaTeX
%% table environment.  Use of deluxetable is preferred.
%%

%% Tables may also be prepared as separate files. See the accompanying
%% sample file table.tex for an example of an external table file.
%% To include an external file in your main document, use the \input
%% command. Uncomment the line below to include table.tex in this
%% sample file. (Note that you will need to comment out the \documentclass,
%% \begin{document}, and \end{document} commands from table.tex if you want
%% to include it in this document.)

%% \input{table}

\clearpage

\begin{deluxetable}{lclclc}
\tablewidth{0pt}
\tablecolumns{6}
\tablecaption{List of molecules used by PHOENIX in E.O.S. and opacity calculations. Number of isotopologues and isotopomers considered for each species included.}
\tablehead{
\colhead{} & \colhead{Number of} & \colhead{} & \colhead{Number of} & \colhead{} & \colhead{Number of} \\
\colhead{Molecule} & \colhead{Isotopologues} & \colhead{Molecule} & \colhead{Isotopologues} & \colhead{Molecule} & \colhead{Isotopologues} \\
\colhead{} & \colhead{\& Isotopomers} & \colhead{} & \colhead{\& Isotopomers} & \colhead{} & \colhead{\& Isotopomers} 

}
\startdata

C$_2$ & 3 & H$_2$O & 4 & NO & 3 \\
C$_2$H$_2$ & 2 & H$_2$O$_2$ & 1 & NO$_2$ & 1 \\
C$_2$H$_6$ & 1 & H$_2$S & 3 & O$_2$ & 3 \\
CH & 2 & H$_3^+$ & 1 & O$_3$ & 3 \\
CH$_3$Cl & 2 & HBr & 2 & OCS & 4 \\
CH$_4$ & 5 & HCN & 3 & OH & 4 \\
CN & 4 & HCl & 2 & PH$_3$ & 1 \\
CO & 7 & HF & 1 & SF$_6$ & 1 \\
CO$_2$ & 8 & HI & 1 & SO$_2$ & 2 \\
COF$_2$ & 1 & HNO$_3$ & 1 & SiH & 3 \\
CaH & 2 & HOCl & 2 & SiO & 4 \\
ClO & 2 & MgH & 3 & TiO & 5 \\
CrH & 1 & N$_2$ & 1 & VO & 1 \\
FeH & 1 & N$_2$O & 5 & YO & 1 \\
H$_2$ & 1 & NH & 2 & ZrO & 7 \\
H$_2$CO & 3 & NH$_3$ & 2 & \nodata & \nodata \\

\enddata
\label{tab:molecules list}
\end{deluxetable}

\clearpage

\begin{deluxetable}{lccrrr}
\tablewidth{0pt}
\tablecolumns{6}
\tablecaption{Comparison of old and new model atoms for atomic species treated 
 in NLTE energy level calculations. Numbers of energy levels and b-b transitions
 given for each ionization stage treated in NLTE. The ground state ionization 
 energy for each species as well as the highest energy levels in the old and new 
 model atoms are also listed.}
\tablehead{
\colhead{} & \colhead{Old Energy Levels} & \colhead{New Energy Levels} &
 \colhead{} & \colhead{} & \colhead{} \\
\colhead{Species} & \colhead{/ Transitions} & \colhead{/ Transitions} &
 \colhead{\raisebox{2pt}{$\chi$}$_{\mathrm{I}}$ (eV)} &
 \colhead{\raisebox{2pt}{$\chi$}$_{\mathrm{High, Old}}$ (eV)} &
 \colhead{\raisebox{2pt}{$\chi$}$_{\mathrm{High, New}}$ (eV)}
}
\startdata

%H I & 80/3160 & 930/1016 & 13.598 & \nodata & 13.596 \\
%He I & 19/37 & 267/605 & 24.587 & \nodata & 24.583 \\
Li I & 57/333 & 60/394 & 5.392 & 5.296 & 5.296 \\
Li II & 55/124 & 55/135 & 75.640 & 74.128 & 74.128 \\
C I & 228/1387 & 230/3262 & 11.260 & 11.155 & 11.155 \\
N I & 252/2313 & 254/3704 & 14.534 & 14.460 & 14.460 \\
O I & 36/66 & 146/855 & 13.618 & 12.728 & 13.482 \\
%Ne I & 26/37 & 262/2437 & 21.565 & \nodata & 21.525 \\
Na I & 53/142 & 58/334 & 5.139 & 5.044 & 5.044 \\
Na II & 35/171 & 35/171 & 47.287 & 45.260 & 45.257 \\
Mg I & 273/835 & 179/1584 & 7.646 & 7.644 & 7.634 \\
Mg II & 72/340 & 74/513 & 15.035 & 14.585 & 14.585 \\
Al I & 111/250 & 115/482 & 5.986 & 5.977 & 5.977 \\
Al II & 188/1674 & 191/2608 & 18.828 & 18.665 & 18.665 \\
P I & 229/903 & 230/945 & 10.487 & 10.266 & 10.266 \\
P II & 89/760 & 90/882 & 19.726 & 17.542 & 17.542 \\
S I & 146/349 & 152/1995 & 10.360 & 10.146 & 10.284 \\
S II & 84/444 & 84/501 & 23.334 & 20.375 & 20.375 \\
K I & 73/210 & 80/576 & 4.341 & 4.300 & 4.300 \\
K II & 22/66 & 22/66 & 31.625 & 27.177 & 27.177 \\
Ca I & 194/1029 & 196/2893 & 6.113 & 6.054 & 6.054 \\
Ca II & 87/455 & 89/760 & 11.872 & 11.641 & 11.641 \\
Ti I & 395/5279 & 555/13304 & 6.820 & 6.653 & 6.653 \\
Ti II & 204/2399 & 204/2586 & 13.577 & 10.504 & 10.504 \\
Mn I & 316/3096 & 297/3067 & 7.435 & 7.418 & 7.418 \\
Mn II & 546/7767 & 512/8299 & 15.640 & 14.968 & 14.968 \\
Fe I & 494/6903 & 902/24395 & 7.871 & 7.539 & 7.815 \\
Fe II & 617/13675 & 894/22453 & 16.183 & 14.665 & 14.814 \\
Co I & 316/4428 & 364/6447 & 7.864 & 7.363 & 7.472 \\
Co II & 255/2725 & 255/2853 & 17.057 & 15.618 & 15.618 \\
Ni I & 153/1690 & 180/2671 & 7.635 & 7.422 & 7.422 \\
Ni II & 429/7445 & 670/17935 & 18.169 & 17.359 & 17.359 \\
\hline \\
Total & 6134/70492 & 8632/130728 & \nodata & \nodata & \nodata \\
\enddata
\label{tab:nlte species list}
\end{deluxetable}

\clearpage

\begin{deluxetable}{lcccccc}

\tablewidth{0pt}
\tablecolumns{7}
\tablecaption{Error estimates in derived LTE ages caused by NLTE effects in IL spectra, and uncertainties in derived ages from our numerical uncertainty and the limiting precision of the Harris Catalog (0.01 dex) \citep{harris96} for all four color indices. Age differences are presented for the bluest and reddest color value for the LTE IL spectra for each index. Subscripts on color indices denote which Set the estimates are associated with.}
\tablehead{
\colhead{Color Index} & \multicolumn{2}{c}{NLTE Effect} & \multicolumn{2}{c}{Numerical} & \multicolumn{2}{c}{Observational} \\
\colhead{} & \multicolumn{2}{c}{} & \multicolumn{2}{c}{Uncertainty} & \multicolumn{2}{c}{Uncertainty} \\
\colhead{} & \colhead{Blue} &  \colhead{Red} & \colhead{Blue} & \colhead{Red} & \colhead{Blue} & \colhead{Red} \\
}

\startdata

$\mathrm{U-B}_1$ & 5.07 & 11.50 & 0.24 & 0.27 & 0.83 & 1.26 \\
$\mathrm{U-B}_2$ & 4.97 & 10.29 & 0.06 & 0.21 & 0.83 & 1.20 \\
$\mathrm{B-V}_1$ & 0.61 & 1.58 & 0.22 & 0.82 & 1.12 & 2.41 \\
$\mathrm{B-V}_2$ & 0.55 & 2.53 & 0.13 & 0.53 & 1.05 & 2.84 \\
$\mathrm{V-I}_1$ & 0.98 & 2.28 & 0.40 & 1.06 & 1.06 & 1.88 \\
$\mathrm{V-I}_2$ & 0.92 & 2.54 & 0.19 & 0.50 & 0.99 & 2.39 \\
$\mathrm{J-K}_1$ & 0.79 & 1.19 & 0.56 & 1.18 & 1.65 & 2.66 \\
$\mathrm{J-K}_2$ & 0.74 & 1.46 & 0.23 & 0.53 & 1.54 & 3.52 \\

\enddata
\label{tab:age errors}
\end{deluxetable}

%% The following command ends your manuscript. LaTeX will ignore any text
%% that appears after it.

\end{document}